\documentclass[12pt,preprint]{aastex}

\newcommand{\be}{\begin{equation}}
\newcommand{\ee}{\end{equation}}
\newcommand{\eint}{{ E_{\rm int} }}
\newcommand{\emix}{{ E_{\rm mix} }}
\newcommand{\excl}{{ E_{\rm excl} }}
\newcommand{\eself}{{ E_{\rm self} }} 

\newcommand{\zhat}{{ {\hat z} }}
\newcommand{\lmech}{{ L_{\rm mech} }}
\newcommand{\fmean}{{ f_{\rm mean} }}  
\newcommand{\porb}{{ P_{\rm orb} }}
\newcommand{\prot}{{ P_{\rm rot} }}
\newcommand{\ergsec}{{ {\rm erg}\,\,{\rm s}^{-1} }}
\newcommand{\asemi}{{ a_{\rm o} }} 
\newcommand{\pmag}{{ {\cal P}_{\rm mag} }} 
\newcommand{\pzero}{{ {\cal P}_0 }}
\newcommand{\pnu}{{ {\cal P}_\nu }}
\newcommand{\bint}{{ B_{\rm int} }} 
 
\newcommand{\ucomp}{{ u_{\gamma} }} 
\newcommand{\tdecay}{{ \tau }}
\newcommand{\nfast}{{ n_{\rm fast} }} 
\newcommand{\euni}{{ |{\bf E}|_{\rm uni} }}
\newcommand{\ecc}{{ \varepsilon }}
\newcommand{\edreicer}{{ |{\bf E}_D| }}

\begin{document} 

\title{\bf MAGNETIC INTERACTIONS IN PRE-MAIN-SEQUENCE BINARIES} 

\author{Fred C. Adams$^{1,2}$, Michael J. Cai$^3$, 
Daniele Galli$^4$, Susana Lizano$^5$, Frank H. Shu$^3$}

\affil{$^1$Michigan Center for Theoretical Physics \\
Physics Department, University of Michigan, Ann Arbor, MI 48109} 

\affil{$^2$Astronomy Department, University of Michigan, Ann Arbor, MI 48109} 

\affil{$^3$Institute of Astronomy and Astrophysics, Academia Sinica, 
Taipei 115, Taiwan, ROC}

\affil{$^4$ INAF-Osservatorio Astrofisico di Arcetri, 
Largo Enrico Fermi 5, I-50125 Firenze, Italy} 

\affil{$^5$Centro de Radioastronom{\'i}a y Astrof{\'i}sica, UNAM, Apartado
Postal 3-72, 58089 Morelia, Michoac\'an, M\'exico}

\begin{abstract} 

Young stars typically have strong magnetic fields, so that the
magnetospheres of newly formed close binaries can interact, dissipate
energy, and produce synchrotron radiation. The V773 Tau A binary
system, a pair of T Tauri stars with a 51 day orbit, displays such a
signature, with peak emission taking place near periastron. This paper
proposes that the observed emission arises from the change in energy
stored in the composite magnetic field of the system. We model the
fields using the leading order (dipole) components and show that this
picture is consistent with current observations. In this model, the
observed radiation accounts for a fraction of the available energy of
interaction between the magnetic fields from the two stars. Assuming
antisymmetry, we compute the interaction energy $\eint$ as a function
of the stellar radii, the stellar magnetic field strengths, the binary
semi-major axis, and orbital eccentricity, all of which can be
measured independently of the synchrotron radiation. The variability
in time and energetics of the synchrotron radiation depend on the
details of the annihilation of magnetic fields through reconnection
events, which generate electric fields that accelerate charged
particles, and how those charged particles, especially fast electrons,
are removed from the interaction region.  However, the major
qualitative features are well described by the background changes in
the global magnetic configuration driven by the orbital motion.  The
theory can be tested by observing a collection of pre-main-sequence
binary systems.

\end {abstract} 


\section{Introduction}

Young stars often have strong magnetic fields on their surfaces, with
typical field strengths $B_\ast\sim 1-2$ kG \citep{jk2009}.  When
binary star systems have sufficiently close separation, the magnetic
fields can interact over the course of the orbit.  For binaries with
circular orbits, variability of the radio emission should not
correlate with orbital phase.  For eccentric binaries, however,
magnetic interactions reach a maximum near periastron and may provide
a means for dissipating energy, a portion of which can be emitted as
synchrotron radiation.

The T Tauri binary V773 Tau A is an observational example of such an
interacting system. The stars have masses $M_1=1.54\pm0.14M_\odot$ and
$M_2=1.33\pm0.097M_\odot$ \citep{boden}. The orbit is observed to have
period $\porb=51.1\pm0.02$ days and eccentricity $\ecc=0.27\pm0.01$. The
estimated stellar radii are $R_1=2.22\pm0.20 R_\odot$ for the primary
and $R_2=1.74\pm0.19 R_\odot$ for the secondary.  This binary system
is actually the ``inner'' binary of a more extended quadruple system;
for purposes of this paper, however, the other two components are too
distant to affect the dynamics and will be ignored.  With the above
orbital parameters, the semi-major axis $\asemi=0.38$ AU.  The
separation $2L$ between the two stars (twice the interaction distance
$L$) equals $\asemi(1+\ecc)\approx0.48$ AU at apoastron and
$\asemi(1-\ecc)\approx0.28$ AU at periastron.  The radio flaring activity
(\citealt{massi2006,loinrod}) varies from a few mJy near apoastron,
where $2L\approx{50}R_\ast$, to a few hundred mJy near periastron,
where $2L\approx{30}R_\ast$, if we adopt a mean stellar radius
$R_\ast= 2 R_\odot$.

\begin{figure} 
\figurenum{1} 
{\centerline{\epsscale{0.90} \plotone{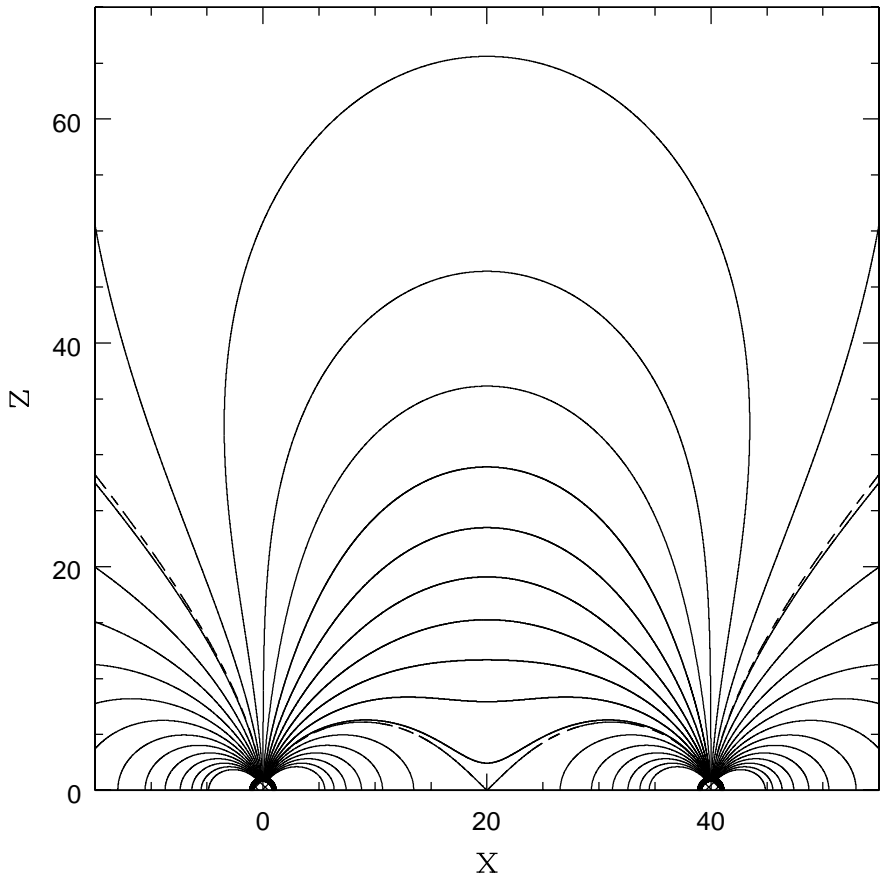} } } 
\figcaption{Magnetic field lines for an interacting binary system 
with separation intermediate between periastron and apoastron (in the
$y = 0$ plane). Lengths are given in units of $R_\ast$.  The field
lines are spaced at uniform intervals on the stellar surfaces. The
stars have the same size $R_\ast$ and surface field strengths
$B_\ast$, and have anti-aligned dipole fields. The dashed curves show
the field lines that mark the boundary between those that close on the
star where they originate and those that extend to the other star.
These limiting field lines effectively loop to the other star via 
$z = \pm \infty$. (Only field lines within the plane containing the 
line of centers of the two stars are shown.) } 
\label{fig:config} 
\end{figure}

\begin{figure} 
\figurenum{2} 
{\centerline{\epsscale{0.90} \plotone{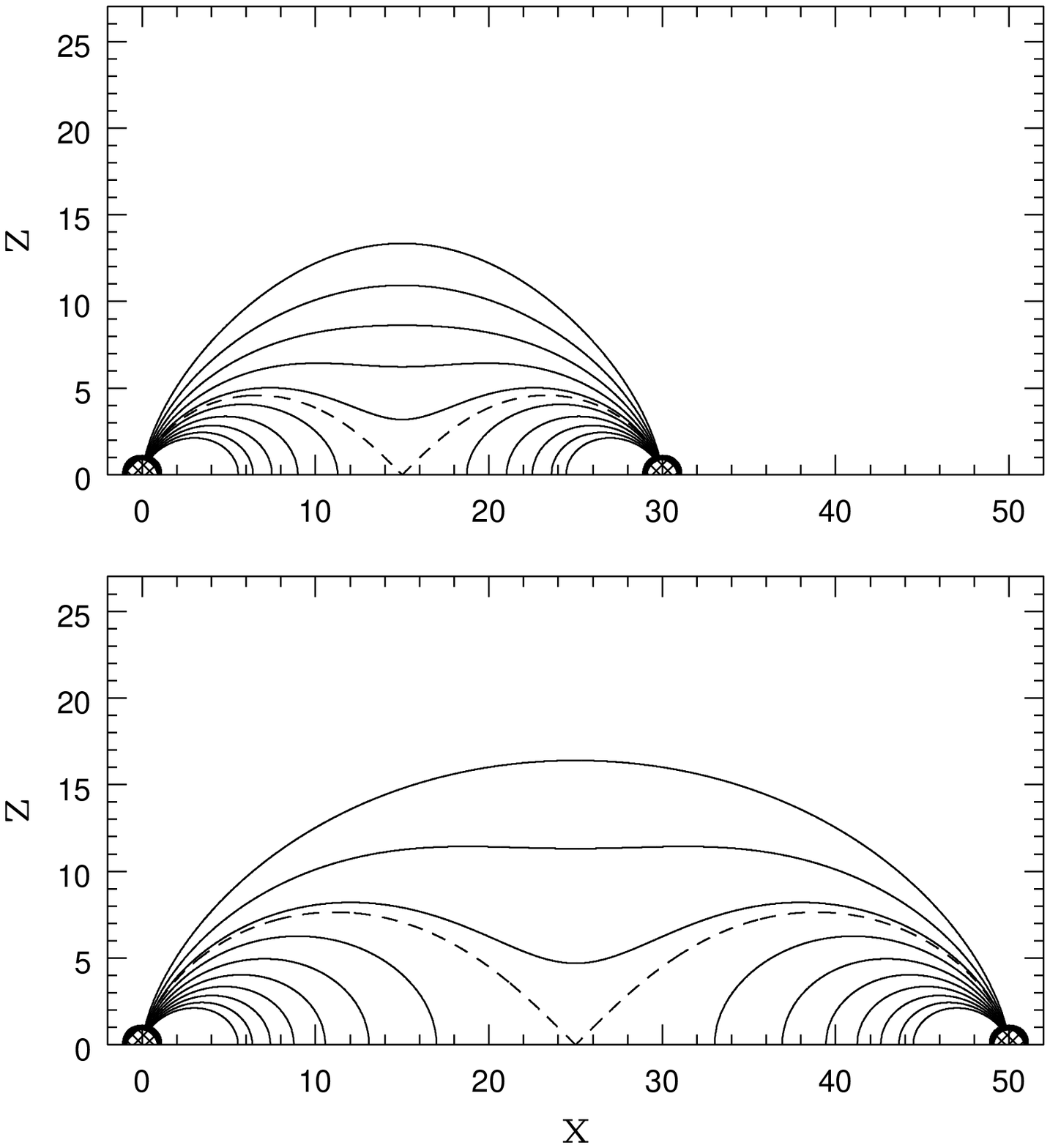} } } 
\figcaption{Magnetic field lines for an interacting binary system at 
periastron (top) and apoastron (bottom) in the $y = 0$ plane (the
plane containing the line of centers of the two stars).  Ten field
lines are shown (solid curves), where the fields lines start at the
same polar angles on the stellar surfaces for both cases.  The stars
have the same size $R_\ast$ and surface field strengths $B_\ast$, and
have anti-aligned dipole fields.  Lengths are given in units of
$R_\ast$.  The figure shows that a larger fraction of the field lines
connect the two stars when they are closer together (compare top and
bottom panels).  The dashed field line (the separatrix) starts at
different angles on the two stars when the system is at apoastron and
periastron. This field line has an X-point, where the field strength B
vanishes. As the stars press together, loops below the X-point
reconnect to become the fields above the X-point. }
\label{fig:blines} 
\end{figure}

The scenario that we envision for this source can be described as
follows: Each member of the binary pair has a strong magnetic field on
its surface. Young stars are often observed to have magnetic field
contributions from several multipoles, with the octupole component
being important near the surface \citep{gregory}; however, the dipole
field dominates at the distance of the interaction region. Over the
binary orbit, the distance between the stars varies in cyclical
fashion. For aligned dipole fields, the magnetic fields become
squeezed together as the stars become closer, and the magnetic field
plays the role of a spring-like restoring force; the gravitational
force is much larger and little energy is dissipated.  For
anti-aligned dipoles, however, the field lines can connect one star to
the other, as depicted in Figure \ref{fig:config}. The field lines
that originate at high latitudes connect one star to the other,
whereas the field lines that originate at low latitudes connect the
star back to itself through loops that cross the equatorial plane on
the sides in the opposite direction from the other star. Note that
these field lines are much like that of a dipole configuration, but
are distorted by the magnetic presence of the companion.  When the
stars are closer, more field lines that start on one stellar surface
end on the other star; the total energy stored in the composite
magnetic field configuration is thus smaller at periastron than for
larger separations. This trend is illustrated by Figure
\ref{fig:blines}, which shows 10 particular field lines when the
system is at periastron (top panel) and apoastron (bottom panel).
Since the stored magnetic energy changes over the orbit, it can
provide power for the observed synchrotron emission.  Note that energy
must be given back to the magnetic field configuration as the stars
move apart --- work must be done on the system. Besides dynamo action
in the stellar interiors, orbital motion provides the ultimate energy
source for pumping up the fields after each dissipation episode. Even
if the energy dissipated in heat/radiation comes entirely from the
orbit, however, the implied changes in the orbital elements are too
small to observe. Although this paper specializes to the case of
anti-aligned dipoles, we note that reconnection is possible for more
general magnetic field configurations (so that the analysis presented
below is qualitatively valid).

Previous models of the radio emission from the V773 Tau A binary
have been proposed.  In particular, the emission could arise from
the interaction of magnetic helmet streamers emanating from the
stellar surfaces (\citealt{massi2006,massi2008}). Our model is
similar in spirit to this previous work, except in our picture, the
helmet streamers are created as a result of the interaction rather
than pre-exist as independent structures that happen to ``bump''
when the stars pass each other at periastron. The scenario explored
herein assumes that the energy comes from the large-scale field
structure (dipoles), whose interaction characteristics are then
computed rather than imposed.

\section{Magnetic Interaction Model}

\subsection{Formulation} 

The two stars of V773 Tau A lack strong infrared excesses, consistent
with their classification as weak-line T Tauri stars. As a first
treatment, we thus assume that the space between the stars contains no
dynamically significant mass from circumstellar disks or their
associated winds (\citealt{kon91,shu94}). Instead, we suppose that any
tenuous plasma that exists above the stellar atmospheres has
negligible inertia relative to the electromagnetic forces. With the
magnetic fields dominant over both matter and gravity, they must
satisfy the force free-condition that any currents present in the
system flow parallel to the lines of force, 
\be\nabla\times{\bf{B}}=\alpha{\bf{B}},\ee 
where $\alpha$ is a function of spatial position.  Nonzero values of
$\alpha$ correspond to non-vanishing electric current. The divergence 
of the above equation, and the condition of no magnetic monopoles
$\nabla\cdot{\bf{B}}=0$, imply that $\alpha$ satisfies the
subsidiary condition ${\bf{B}}\cdot\nabla\alpha=0$, i.e., $\alpha$ 
is constant along field lines. 

Suppose current does not flow to/from infinity, i.e., the system
neither gains nor loses net electric charge over time.  Then
$\alpha=0$ on open field lines and hence the open field lines must
satisfy the vacuum-field equations. The closed field lines have two
topological types: [1] The field lines emerge from the surface of a
star and submerge beneath the surface elsewhere on the same star.  [2]
The field lines emanate from the surface of one star and terminate
beneath the surface of the second star.  In either case, in the
presence of finite resistivity, any currents will dissipate with time
and will asymptotically approach zero unless dynamo action restores
the differential stresses on electrons and ions that led to the
currents in the first place.  Because current dissipation involves the
transformation of field energy into heat and/or radiation (e.g.,
synchrotron emission if non-thermal processes of accelerating charges
are involved), vacuum field configurations represent a lower state of
magnetic energy than their force-free counterparts.  The important
point is that current dissipation and magnetic-field reconnection tend
to enforce $\alpha\rightarrow0$, i.e., the closed field lines approach
a vacuum-field configuration. For the sake of completeness we note
that a pure vacuum field configuration may not be fully accessible,
since finite resistivity only pushes the system towards the lowest
energy state consistent with its helicity. Here we assume that the
helicity is small and consider vacuum-field configurations for both
open and closed field lines defined by the equations
\be\nabla\times{\bf{B}}=0\qquad{\rm and}
\qquad\nabla\cdot{\bf{B}}= 0.\label{vacuumfield}\ee 
We are thus computing the minimum energy states that would be present
in the magnetically interacting binary system if current dissipation
occurs on a timescale rapid compared to the orbital period.  This
philosophy informs the rest of our analysis until we get to \S 2.4.

\subsection{Energy of External Magnetic Fields} 

As a starting approximation, we take all field lines to be closed, so
that the magnetic field strength vanishes at infinity. In this case,
the most general field configuration associated with either star is an
exterior multipole expansion. A magnetic multipole of order $\ell\ge1$
decreases with distance $r_j\equiv|{\bf r}-{\bf x}_j|$ from the
stellar center as $r_j^{-(\ell+2)}$, where ${\bf x}_j$ is the stellar
position ($j=1,2$).  In the V773 Tau A system, magnetic interactions
occur at distances that are large compared to the stellar radii. To
leading order, we can ignore all multipoles higher than the dipole
component $\ell = 1$. The vacuum field configuration of the two stars
can then be written
\be{\bf{B}}={\bf{B}}_1+{\bf{B}}_2,\ee
where both stellar fields are pure dipoles with moments 
${\bf m}_j=m_j {\hat z}$, 
\be{\bf{B}}_j={{3({\bf m}_j\cdot{\bf r}_j){\bf r}_j\over r_j^5}}-{{\bf m}_j\over r_j^3} 
= {m_j \over r_j^3} \left(2\cos\theta\,{\hat r}_j
+\sin\theta\,{\hat\theta}_j\right)\,,\ee
where ${\bf r}_j={\bf r}-{\bf x}_j$ and $r_j = |{\bf r}_j|$. 
Note that this form for the dipole field and the corresponding
definitions of the magnetic moments follow Jackson (1962); in
particular, $m_j = B_j R_j^3$, where $B_j$ is the equatorial magnetic
field strength on the stellar surface and $R_j$ is the stellar radius 
(for $j$ = 1,2).

Let $E_B$ be the energy in the magnetic field exterior to the
two stars. If $V$ denotes the volume of all space excluding the two
spheres instantaneously occupied by the stars, $E_B$ is given by
\be{E_B}={\int_V}{B^2\over 8\pi}d^3r=
{\int_V}{B_1^2\over 8\pi}d^3r+\int_V {B_2^2\over 8\pi}d^3r+
{\int_V}{{\bf{B}}_1\cdot{\bf{B}}_2\over4\pi}d^3r.\ee 
For each star, let $V_j$ denote the entire space excluding the volume
occupied by star $j$ at its instantaneous position. The total magnetic
energy $E_B$ can be separated,
\be{E_B}=\eself+\eint,\ee
where $\eself$ represents the magnetic field energy that applies 
for isolated stars, 
\be{\eself}=\int_{V_1}{B_1^2\over 8\pi}d^3r+\int_{V_2}{B_2^2\over 8\pi}d^3r,\ee
and where $\eint$ represents the difference in magnetic field
energy resulting from interaction of the stellar fields,
\be{\eint}=\int_V{{\bf{B}}_1\cdot {\bf{B}}_2\over 4\pi}- 
\int_{V_2^*}{B_1^2\over 8\pi} d^3r-\int_{V_1^*}{B_2^2\over 8\pi} d^3r\label{eintint}\ee
where the $V_j^*$ are the spherical volumes of the stars.  

Independent of the stellar motion, $\eself$ is constant provided that 
internal magnetohydrodynamics can maintain constant magnetic dipole
moments ${\bf m}_j$ within each star. This self energy $\eself$
provides a benchmark energy 
\be{\eself}=E_1+E_2,\ee 
where 
\be{E_j}={B_j^2{R_j^3}\over 4}\int_{-1}^{1}d\mu\int_1^\infty\xi^2{d\xi}\left[1+3\mu^2\right] 
\xi^{-6}={B_j^2R_j^3\over3},
\ee 
where the dimensionless variables are $\mu=\cos\theta$ and
$\xi=r/R_j$.  For the system parameters adopted below, $\eself$ =
$4.1\times 10^{39}$ erg.

The interaction energy $\eint$ varies over the orbit and contributes
to the observed variable radio emission, provided that the orbit is
eccentric so that the separation $2L$ changes with time.  The final
two terms in equation (\ref{eintint}) for $\eint$ involve the negative
field energies of the individual stars in the excluded volumes of the
other star; we called the sum of these two terms $E_{\rm excl}$.  
The first term involves the mixed dot product, which can
be positive or negative, depending on whether the magnetic fields from
the two stars in the interaction region are primarily aligned
(${\bf{B}}_1\cdot{\bf{B}}_2>0$) or anti-aligned
(${\bf{B}}_1\cdot{\bf{B}}_2<0$); we call this term $E_{\rm mix}$. 
The maximally negative interaction energy $\eint=\excl+\emix$ arises
when the stellar dipoles are anti-aligned: In this case, oppositely
directed field lines can squeeze together, reconnect, and release
energy (ultimately to accelerate particles that create radiation).
For the aligned case, the fields would dissipate at a (much) lower
rate because the current density is lower.

Here we specialize to the antisymmetric case, where the stars have
identical radii $R_1 = R_2 = R_\ast$ and magnetic dipole moments of
identical strength that are parallel and antiparallel to $\zhat$, the
unit normal of the orbital plane, i.e., ${\bf m}_1=B_*R_*^3\zhat$ =
$-{\bf m}_2$. For the sake of definiteness, we adopt a standard value
$B_\ast$ = 1.5 kG for both stars.  Although Johns-Krull (2009) quotes
an average magnetic field strength in T Tauri stars of 2.5 kG, this
value includes all multipoles, so it still might be overly-optimistic
to attribute $B_\ast = 1.5$ kG to the dipole component. However, both
stars in the V773 Tau A system are K stars with deep outer convection
zones, and have short rotation periods under 3 days (Welty 1995),
properties that correlate with stronger surface field strengths.  As a
result, the V773 stars, which may well be pre-main-sequence
progenitors of RS CnV stars, could be expected to support unusually
well-ordered and coherent (although variable) fields.

We define $2L$ to be the instantaneous center-to-center distance
between the stars.  The midpoint between the stars defines the origin
of the coordinates. For stars of unequal masses, the origin is not at
the center of mass and the frame of reference is accelerating rather
than inertial.  Moreover, the coordinate axes rotate at an
instantaneous angular velocity $\Omega_{\rm cm}$ as the stars orbit
about their center of mass. Given the rotation periods under 3 days,
the spin angular velocities $\omega_j$ of both stars exceed
$\Omega_{\rm cm}$ even at periastron.  Thus, the system will probably
support a unipolar inductor with a form reminiscent of the Io-Jupiter
system \citep{goldbell}.  The induced electric field could accelerate
fast particles that augment the radio emission, and this process would
have a periodicity correlated with the orbital phase. However, the
changing magnetic field configuration, considered here, provides a
larger induced electric field (see Section 2.4).  As a result, we
leave the unipolar inductor for future work and focus on the simpler
quasi-magnetostatic concepts that we believe are responsible for the
radio synchrotron emission.

First we estimate the mixed dot-product term $\emix$ as follows: For
identical stars with anti-aligned magnetic dipoles, one can show that
the composite magnetic field lines are symmetric with respect to the
mid-plane perpendicular to the line connecting the stellar centers;
this plane lies a distance $L$ from either star. We assume that the
interaction energy is given by the magnetic energy density integrated
from this plane outward to infinity, i.e., the magnetic energy in this
volume is lost because the fields from the two stars are connected.
Introducing a form factor $f(\varpi,\varphi,z)$ to take care of the geometric
vagaries of the dot product of the mixed fields, $\emix$ can be
written
\be\emix=2 \int_0^{2\pi} d\varphi  \int_L^\infty dz\int_0^\infty\varpi{d\varpi} 
{B_\ast^2{R_\ast^6}\over{8\pi}}{f(\varpi,\varphi,z)\over(\varpi^2+z^2)^3} 
=\fmean\left({B_\ast^2{R_\ast^6}\over{24L^3}}\right),\label{eintest}\ee
where we have used a cylindrical coordinate system with the $z$ axis
pointing along the line connecting the stars; the mean-value theorem
allows us to take $f(\varpi,\varphi,z)$ out of the integral and replace it
with an appropriate mean value $\fmean$.

Next we estimate the energy from the excluded volume integrals by
replacing the field strength at each point in the volume by its value
at the center of the sphere and then multiplying this value by the
volume of the sphere.  If we introduce a factor $f_{\rm excl}$ to make
the result exact, two such integrals yield
\be\excl=f_{\rm excl}\left({B_\ast^2R_\ast^9\over{192}L^6}\right).\label{exclude}\ee
We expect $f_{\rm excl}\sim{1}$ when $2L\gg{R_*}$, which holds for the
V773 Tau A system. However, the excluded volume energy $\excl$ is
smaller in magnitude than the contribution $\emix$ by a factor of
$\sim8(L/R_\ast)^3\sim10^5$ (for V773 Tau A system parameters). The
excluded volume energy can thus be neglected to leading order.

Using the above estimates, the interaction energy is given numerically by
\be
\eint\approx \fmean\left(8.8\times10^{34}{\rm erg}\right)\, 
\left({B_\ast\over 1500 {\rm G}}\right)^2
\left({R_\ast\over{1.4\times10^{11}}{\rm cm}}\right)^6\, 
\left({L\over{2\times10^{12}}{\rm cm}}\right)^{-3}.
\label{eintnum}
\ee
Since $\fmean$ is positive for the parallel case and negative
for the anti-parallel case, we expect significant release of magnetic
energy (yielding enhanced synchrotron radiation) only when the
magnetic dipole moments of the two stars are (mostly) anti-aligned.

\begin{figure} 
\figurenum{3} 
{\centerline{\epsscale{0.90} \plotone{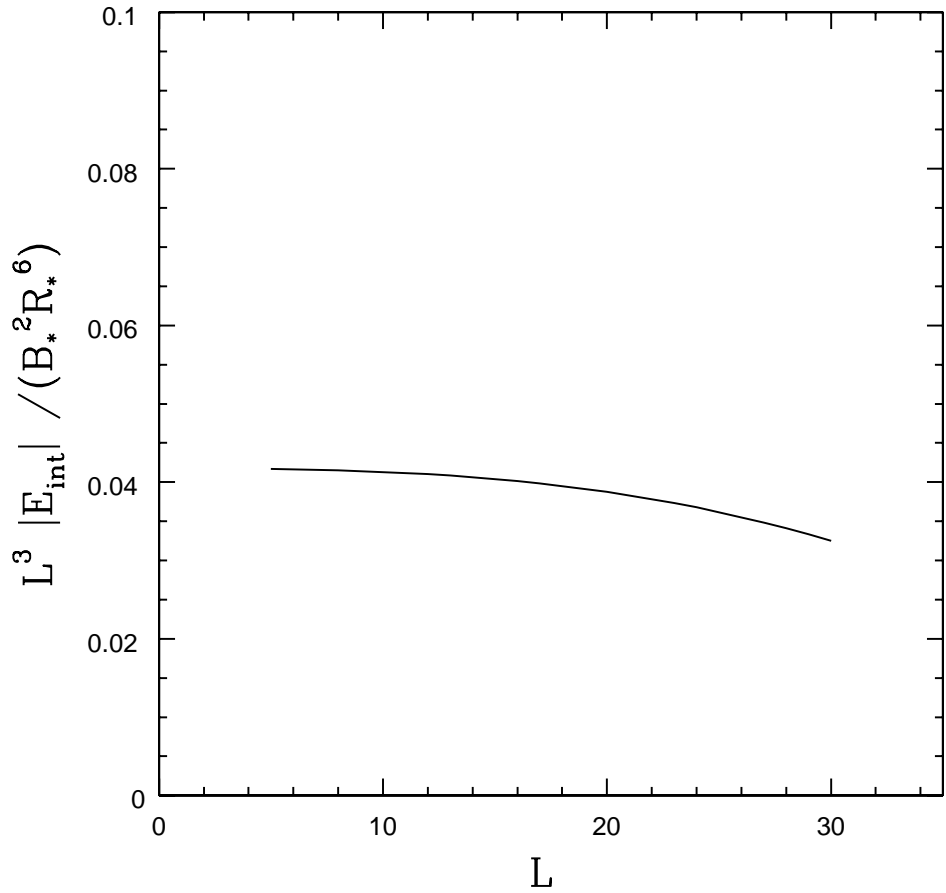} } } 
\figcaption{Dimensionless interaction energy versus stellar 
half-separation, where $L$ is measured in stellar radii and the energy
is scaled by $L^3$. The stars have the same radii and surface field
strengths, and have anti-aligned dipole magnetic fields. The scaled
interaction energy $L^3\eint$ varies slowly with $L$, so that
$\eint\sim{L^{-3}}$. }
\label{fig:eintvsa} 
\end{figure}

Next we determine the magnetic interaction energy by numerically
evaluating the integral for the mixed dot-product term in equation
(\ref{eintint}).  Figure \ref{fig:eintvsa} plots the resulting
dimensionless interaction energy
$-\eint{L^3}/(B_\ast^2R_\ast^6)\approx-\fmean/24$ versus the
half-distance $L$ between the stars. Since this quantity has little
variation with the binary separation $2L$, the interaction energy
scales as $\eint \propto {L^{-3}}$ to good approximation (consistent with
the estimate from equation [\ref{eintest}]). Further, the inferred
value of $\fmean$ is close to unity. As a result, the maximum energy
released over the orbital period of the V773 Tau A binary is
\be
E_{\rm int}^{\rm apo} - E_{\rm int}^{\rm peri} \sim  
6.1 \times 10^{34}\; {\rm erg}.
\label{apoperi}
\ee
Notice that the energy difference of $\eint$ between apoastron and
periastron --- that available to convert into radiation --- is only a
small fraction of $\eself$ (about one part in $10^5$).  Nevertheless,
this energy difference is sufficient to explain the observed
synchrotron emission.

\begin{figure} 
\figurenum{4} 
{\centerline{\epsscale{0.90} \plotone{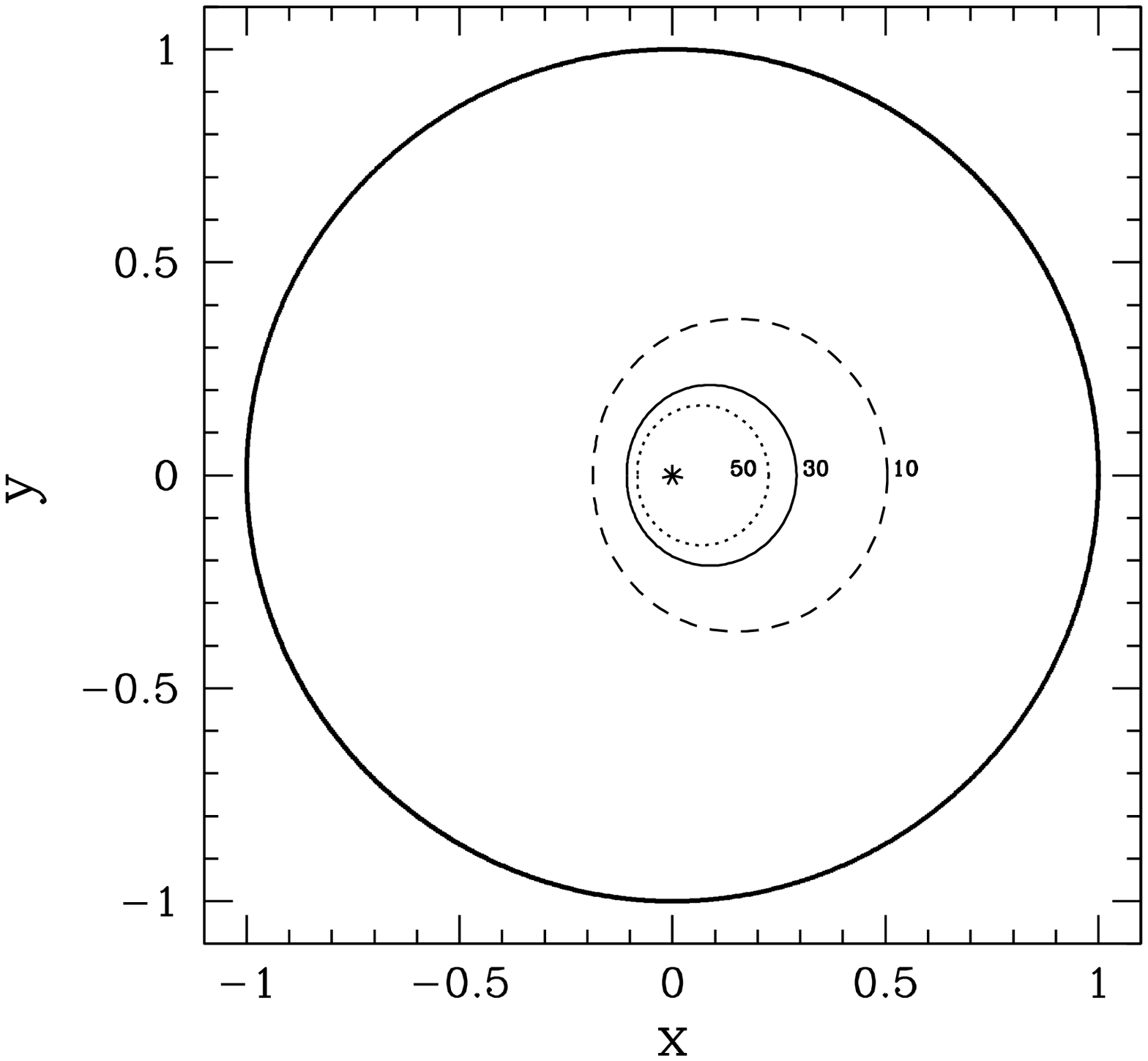} } } 
\figcaption{Polar cap region for one star of an interacting binary 
for varying separations. The companion star is to the right of the
figure.  The polar cap is the area of the star for which magnetic
field lines that originate on one stellar surface terminate on the
other stellar surface. The contours shown here are projections in the
$x-y$ plane. There are actually two projections, one for field lines
above $z = 0$ (the orbital plane) and one for below $z = 0$ that lie
on top of each other in projection.  Polar cap regions are shown for
stellar separations $2L/R_\ast=10$ (dashes), $30$ (solid), and $50$
(dots).  The heavy circle depicts the stellar equator.} 
\label{fig:polarcap} 
\end{figure}

As energy stored in the magnetic field configuration varies over the
binary orbit, magnetic field lines must change their form.  They
become increasingly ``connected'' as the stars become closer (see
Figure \ref{fig:blines}).  The field lines that begin on one stellar
surface and end on the other stellar surface generally originate near
the stellar poles. The stars thus have a ``polar cap'' region where
shared field lines begin. As the stars move farther apart (closer
together), the size of the polar cap decreases (increases).  Figure
\ref{fig:polarcap} illustrates this behavior by showing the projected
polar cap for varying stellar separations.  The boundaries of these
polar regions are obtained by numerically integrating the magnetic
field lines outward from one stellar surface to determine if the field
line intersects the other stellar surface or returns to the original 
one. Note that the polar caps are not symmetric with respect to the
geometrical pole of the star (that defined by the $\zhat$-axis), but
rather are ``tilted'' toward the other star (which lies far to the
right in the figure).  More specifically, for each value of $y$ on the
polar cap contour, there are two values of $x$.  The field line that
attaches to the larger algebraic value of $x$ connects to the other
star through the analogue of the X-point (drawn in Figures
\ref{fig:config} and \ref{fig:blines} for $y = 0$) in the equatorial
plane (also the plane of the binary orbit).  The field line that
attaches to the smaller algebraic value of $x$ connects to the other
star through infinity (the limiting field line on the back side of the
star in Figure \ref{fig:config} for $y$ = 0, i.e., the dashed curve
that arches higher and higher in $z$). Finally, we note that Figure
\ref{fig:polarcap} shows that the solid angle $A_c$ of the polar cap
is given by the approximate expression $A_c\approx 2R_\ast/L$ (see
Appendix \ref{sec:polarcap} for a derivation).

\subsection{Power from Magnetic Interactions}  

The above discussion estimates the magnetic interaction energy
available to convert into synchrotron radiation as a function of
stellar separation. The corresponding power $\pmag$ available for
particle acceleration (and radiation) is the negative time-rate of
change of the magnetic energy and can be written 
\be{\pmag}=-{d\eint\over{dt}}=3\eint{1\over{r}}{dr\over{dt}},\label{pzero}\ee
where $r$=$2L$ is the instantaneous distance between the stars. 
The stars execute a Keplerian orbit, 
\be{r}={\asemi(1-\ecc^2)\over{1+\ecc\cos\theta}},\label{kepler}\ee
with semi-major axis $\asemi$, eccentricity $\ecc$, and orbital angle
$\theta$ \citep{md99}. Periastron occurs when $\cos\theta$=1.
The orbital angular speed follows Kepler's second law, 
${\dot \theta} = J/r^2$, where the specific angular momentum
$J={\asemi^2\,\Omega(1-\ecc^2)^{1/2}}$, and $\Omega=2\pi/\porb$ 
is the mean angular velocity of the orbit ($\porb\approx51$ days).
Collecting the above results, the power becomes
\be{\pmag}= f_{\rm mean}{\Omega{B_\ast^2}R_\ast^6\over{\asemi^3}}
\left({1+\ecc\cos\theta\over{1-\ecc^2}}\right)^4
{\ecc\sin\theta\over(1-\ecc^2)^{1/2}}.\label{power}\ee
The positive sign in equation (\ref{power}) indicates that magnetic
energy is released as the stars become closer together.

\begin{figure} 
\figurenum{5} 
{\centerline{\epsscale{0.90} \plotone{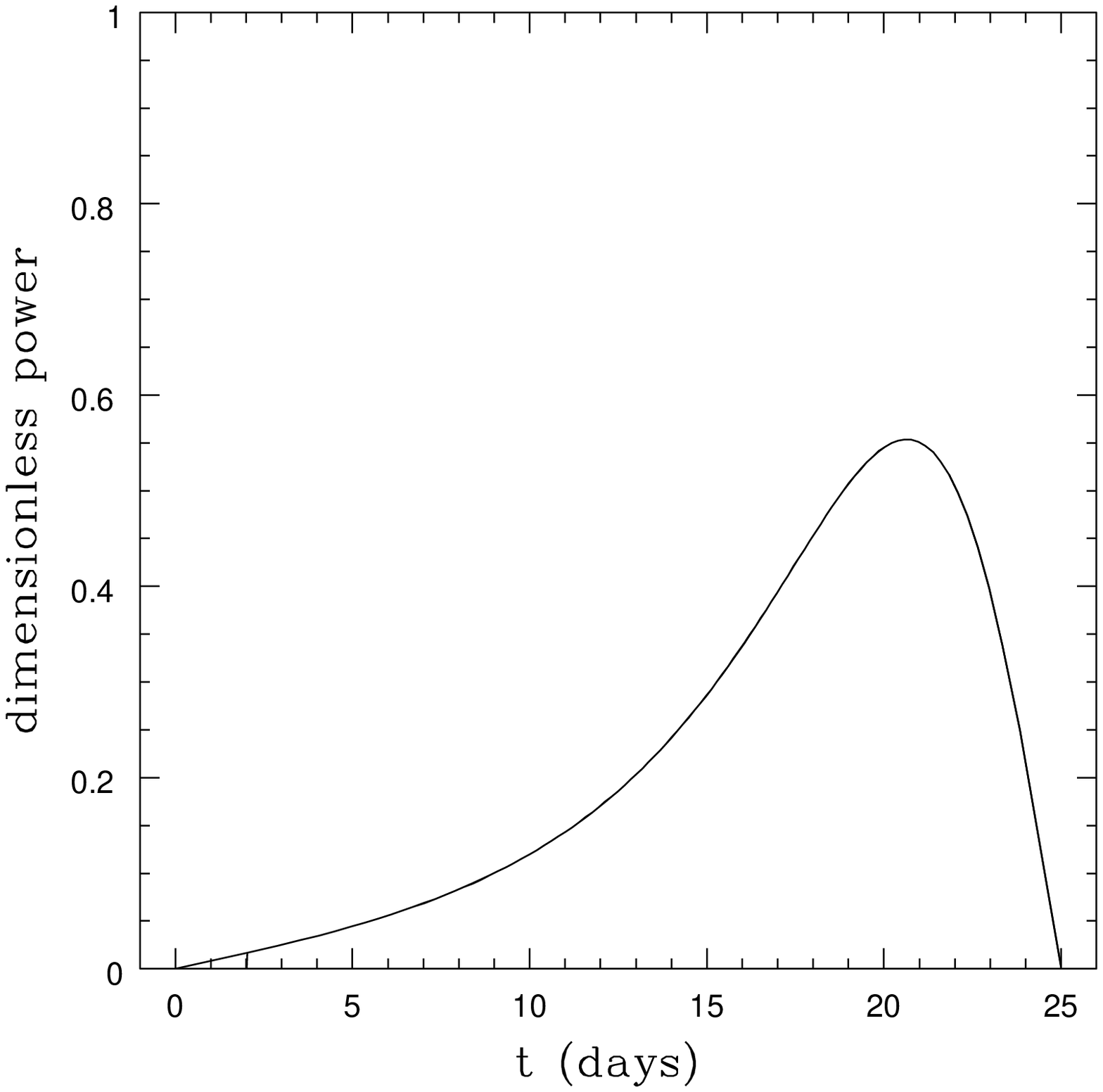} } } 
\figcaption{Power dissipated by the magnetic field configuration 
of the binary system over one-half orbit. Time is measured in days
since apoastron.  The dimensionless power $|\pmag/\pzero|$ is plotted
on the ordinate, where the fiducial power scale $\pzero$ is defined by
$\pzero=B_\ast^2R_\ast^6\Omega/\asemi^3 \approx 1.3\times10^{29}$
$\ergsec$ (see equation [\ref{powerscale}]).  Energy is dissipated
over the incoming portion of the orbit. }
\label{fig:power} 
\end{figure}

Figure \ref{fig:power} shows the magnitude of the total power as a
function of time, which is measured in days since apoastron. The plot
presents the dimensionless power $|\pmag/\pzero|$, where we have defined
the fiducial power scale
\be
{\pzero}= {B_\ast^2R_\ast^6\Omega\over\asemi^3}
\approx 1.3\times10^{29}\,\ergsec.
\label{powerscale}
\ee
Note that the power vanishes at periastron when $\sin\theta=0$,
whereas peak power occurs $\sim4$ days earlier.  For simplicity, 
in the figure we assumed $|f_{\rm mean}|=1$. 

The synchrotron emission is not necessarily fully correlated with the
magnetic dissipation: Although the {\it deformation} of the magnetic
field is slow and steady, with timescales characteristic of periastron
passage (i.e., days; see Fig. \ref{fig:power}), the {\it energy}
{\it release} from the magnetic fields and the resulting radio flares
can take place on significantly shorter timescales. For example, the
Sun twists up its surface fields on a natural timescale given by the
differential rotation between its equator and pole (many weeks); when
outbursts arise, however, they are much more rapid, lasting only
minutes for impulsive flares and hours for coronal mass
ejections. This phenomenon is much like the buckling of a metal
plate stressed along its edges: the accumulation of stresses can
be slow, but the relief of those stresses --- after they become
supercritical --- can occur on catastrophically fast timescales.

\subsection{Acceleration of Particles to Relativistic Energies}

The loss of magnetic field energy does not directly result in
radiation because the orbital frequencies, or even peak rates of
reconnection, are too slow to result in meaningful electromagnetic
radiation.  Instead, when magnetic energies are changing because
of annihilation of magnetic field ${\bf B}$, electric fields 
${\bf E}$ arise that are given by Faraday's law of induction:
\be
\nabla \times {\bf E} = - {1 \over c} 
{\partial {\bf B} \over \partial t},
\ee
with $c$ equal to the speed of light in a vacuum.  Although the
resulting electric field is easy to compute as a function of space
and time in the vacuum-field approximation, we do not record the
results here.  We merely comment that this electric field $\bf E$
does have a component parallel to the magnetic field $\bf B$, and
that ${\bf B}\cdot {\bf E}$ along magnetic field lines changes sign
over the orbit cycle, so that the electric currents that arise in
response to the electromagnetic fields are alternating in character
and do not involve a permanent charging up of either star.

The resulting electric field ${\bf E}$ that is not strictly
perpendicular to the magnetic field ${\bf B}$ then accelerates
particles, in particular electrons, and it is these fast electrons, if
they achieve relativistic energies, that emit synchrotron radiation in
the ambient magnetic field. Some of the fast electrons may have mirror
points below the surface of one of the stars in the system. In that
case, electrons can spiral to the footpoints of the magnetic field and
collide with ions in the dense atmosphere of the polar cap of that
star.  Such collisions produce brehmsstrahlung radiation, which
provides another channel through which the loss of magnetic energy can
appear as electromagnetic radiation.

\subsection{The Source Region for Fast Electrons}

In a fully ionized plasma, such as may exist in the upper
chromospheric regions of the two weak-lined T Tauri stars, the
effective frequency at which electrons are slowed down by Coulomb
collisions with other charged particles decreases with increasing
energy $\gamma m_e c^2$, where $m_e$ is the rest mass of the electron
and $\gamma = (1-v^2/c^2)^{-1/2}$ is its Lorentz factor, and where
${\bf v}$ is the electron velocity.  As a consequence, in the presence
of a non-vanishing component $E_\parallel$ of the mean electric field
${\bf E}$ that is parallel to the mean magnetic field ${\bf B}$
(i.e., $E_\parallel\equiv{\bf E}\cdot{\hat b}$ where ${\hat b}$ = 
${\bf B}/|{\bf B}|$) electrons above a certain threshold energy can
run away to ever higher energies (Dreicer 1959):
\be
m_e c^2 \left( {d \gamma \over d t} \right)_{\rm elec} =
-e {\bf E} \cdot {\bf v},
\label{accel}
\ee
where $e$ is the charge of the electron and the important component
of ${\bf v}$ in this context is that parallel to ${\bf B}$.

The ratio of the magnitude of the quasi-electrostatic force to the
gravitational force at the poles of either star of a particle with
charge $e$ and mass $m_e$ is $e|E_\parallel|R_*^2/GM_*m_e$ = 
$\Gamma |E_\parallel/B_*|$ where $\Gamma \equiv eB_*R_*^2/GM_*m_e \sim
10^{20}$. Thus, when $|E_\parallel|$ is larger than $\sim10^{-17}B_*$,
which is true of the vacuum field values, the electrostatic
acceleration of electrons (which drag ions with them) off of the poles
of one star (or the other) will more than overcome the gravitational
attraction of the stars for the plasma.

Space charge can accumulate and AC current flow can occur that modify
the self-consistent fields relative to the vacuum approximation of
two (naked) moving magnetic dipoles.  The synchrotron radiation
observed in the system suggests that there are $n_e \sim 2 \times
10^5$ electrons per cm$^3$ in a volume of space $L^3$ when the system
is at periastron $L = 15 R_*$.  If these electrons all came from two
polar caps, each of area $8 \pi R_*^3/L$ supplied at a current flow
$I \sim (16 \pi R_*^3/L) e n_{e*}c$ over a time equal to half an
orbital period $\porb/2$, electron number conservation requires
$I \porb/2 = en_e L^3$.  In other words, the number density of
fast electrons in the source region of the polar caps must be
$n_{e*} \sim n_e (L^4/8\pi R_*^3 c \porb) \sim 426$ cm$^{-3}$. 
The magnetic field associated with the current flow $B_{AC} \sim
e n_{e*}R_*/4\pi \sim 2300$ G is comparable to the strength of the
dipole magnetic field $2B_* = 3000$ G at the stellar pole.  Thus, we
expect the back reaction of the induced current flow of fast
electrons to interfere by an order unity amount with the vacuum
approximation.

An important quantity in this problem is the number density $\nfast$
of fast (runaway) electrons that develop when the thermal plasma of a
local number density $n_e$ in the chromosphere or corona of one of the
stars is subject to acceleration by an electric field $E_\parallel$ of
appropriate sign parallel to the polar-cap magnetic field of strength
$2B_*$. Following the treatment of Tandberg-Hanssen \& Emslie (2009),
we can write the rate of fast electron production in the form
\be
{d \nfast \over dt} = 0.35 n_e \nu_c f(\epsilon) \, ,
\ee  
where $\epsilon=|E_\parallel|/\edreicer$, the strength 
$\edreicer$ of the Dreicer electric field is given by 
\be
\edreicer = {e \over \lambda_D^2 } \ln \Lambda \, ,
\ee
and where this treatment is valid in the limit $\epsilon \ll 1$. 
The parameter $\lambda_D=(kT/4 \pi n_e e^2)^{1/2}$ is the Debye length
and the Coulomb logarthim $\Lambda$ takes the form $\ln \Lambda$ = 
$18 + \ln(3.16 \times 10^{-5} T^{3/2} n_e^{-1/2})$.  Finally, 
$\nu_c$ is the electron-ion collision-frequency,
\be
\nu_c = 3.6 \, T^{-3/2} n_e \ln \Lambda \, , 
\ee
where all quantities are in cgs units. In the regime $\epsilon<1$, 
the function $f(\epsilon)$ has a simple form (Tandberg-Hanssen \&
Emslie 2009); however, for the typical parameters of this problem
($n_e \sim 4 \times 10^5$ cm$^{-3}$ and $T\sim 2 \times 10^7$ K), 
the Dreicer field $\edreicer \sim 2 \times 10^{-11}$ statvolt/cm, so
that $\epsilon \gg 1$. In this regime almost all of the electrons are
subject to runaway.

The resulting flow of current, which is almost instantaneous on the
orbital time scale, will charge up both stars (one negatively and one
positively), and nothing in between, because the induction electric
field satisfies $\nabla\cdot {\bf E} = 0$.  The electrostatic
potential associated with the negative charge grows to provide just
enough repulsion of electrons (with a similar development for the
positive charge on the other star in repelling ions) so that the AC
current automatically shunts back and forth to maintain a quasi-steady
state.  In other words, a large-scale dipole electrostatic field (with
one pole on each star) develops so that it compensates macroscopically
for the induced electric field of the two moving magnetic dipoles.
Because of the different spatial dependence of the fields of two
moving electric monopoles and two moving magnetic dipoles, the
cancellation of electric fields only holds as an average for the polar
caps and not throughout space.  Acceleration of electric charges in
between the two stars occurs in response to the superposed magnetic
and electric fields.  In what follows, we provide order of magnitude
estimates by considering only the action of the induction electric
field. 

\subsection{Constraints Set by the Observed Synchrotron Radiation}

To make a numerical estimate for the electron acceleration that occurs
in the spaces between the two stars after the electrons leave the
source and sink regions near the stellar poles, we need to specify a
``typical'' field strength $B_{\rm int}$ that is undergoing change on
a time scale $t_B$ over a length scale $L$ in the interaction region.
As a typical $B_{\rm int}$, we therefore take the average of its value
at the X-point (where $B$ = 0) and its value halfway between one of
the stars and the X-point. On a direct line, this means that the
distance from one star is $L/2$ and from the other star, $3L/2$.  For
anti-aligned dipoles, the magnetic field strength at the latter point
equals 
\be
B_{1/2} = B_\ast R_\ast^3
\left[ \left({1\over L/2}\right)^3 - \left(1\over 3L/2
\right)^3\right]
= {208\over 27}B_\ast\left({R_\ast\over L}\right)^3.
\label{bhalf}
\ee
For our standard parameters, $B_{1/2}\approx3.96$ G.  An average of
this value and the X-point value ($B_X=0$) produces the estimate
\be
B_{\rm int} = 2 \; {\rm G}.
\ee
The magnitude of the electric field is given by
\be
\left | {\bf E}\right | \sim 
\left({L \over c \ \tdecay}\right) \bint ,
\label{estfield}
\ee
where $\tdecay$ is the time scale on which the magnetic field
changes.  This time scale is bounded from above by the orbit time
$\porb \approx 4.4 \times 10^6$ s and is bounded from below by the
burst time scale $t_B = 10^4$ s (from the observations of Massi et
al. 2006, 2008). For all possible values of the time scale $\tdecay$,
the length scale $c\,\tdecay \gg L = 2 \times 10^{12}$ cm and hence
$E\ll{B_{\rm int}}$ in cgs units.  Using a (conservative) value near
the upper end of the range, $\tdecay = 10^6$ s, we find that
$c\,\tdecay=3\times10^{16}$ cm and $|{\bf E}|\sim10^{-4}B_{\rm int}$.
Nevertheless, as we shall see, the action of this electric field over
the long distances available in the system can achieve impressive
accelerations of charged particles.
 
In this work we assume that the electric field ${\bf E}$ generated
through Faraday induction (from equation [\ref{estfield}]) is the
dominant contribution. In particular, we verify below that $|{\bf E}|$
is larger than the field of a unipolar inductor, where an electric
field is produced by the slippage of magnetic fields through the
stellar atmosphere(s). This latter field strength, denoted here as
$\euni$, can be estimated through the expression
\be
\euni \sim {1 \over c} \left|{{\bf v} \times {\bf B}}\right|
\sim {2 \pi R_\ast \over c \prot} B_\ast
\left({ R_\ast \over 2 L} \right)^3 \, ,
\ee
where we estimate the speed $v=|{\bf v}|$ of slippage as
\be
v \sim {\cal F} \, {2 \pi R_\ast \over \prot} \, ,
\ee
where $\prot$ is the rotation period of the star, and where the
dimensionless fraction ${\cal F} \sim 1$ because the orbital angular
speed of the companion star is small compared to rotation speed of the
star itself.  Note that to obtain this estimate, we evaluate the
magnetic field strength from one star at the position of the other
star. Comparing this result to the electric field obtained from
Faraday induction via equations (\ref{bhalf} -- \ref{estfield}), we
find the ratio
\be
{|{\bf E}| \over \euni } = {416 \over 27 \pi {\cal F}}
{L \over R_\ast} {\prot \over \tdecay} \sim 20 \, ,
\ee
where we have used $\tdecay=10^6$ s and ${\cal F}=1$. The electric
field from Faraday induction (used here) is thus safely larger than
that of the unipolar inductor, although the latter could be included
as a correction in future work.

The maximum frequency of the synchrotron radiation produced by an
electron with Lorentz factor $\gamma$ spiraling in a magnetic field
of strength $B$ (see, e.g., Shu 1991) is given by
\be
\nu = \gamma^2 \nu_L \qquad {\rm where} \qquad
\nu_L = {e B \over 2 \pi m_e c} \, ,
\ee
i.e., $\nu_L$ is the Larmor frequency.  For $B = \bint = 2$ G, in
order to produce $\nu=90$ GHz radiation, $\gamma$ must be at least
$127$.  Such a Lorentz factor corresponds to moderately relativistic
electrons, and we may approximate $v\approx{c}$ when $v$ stands alone,
such as in equation (\ref{accel}).  We now consider the length of time
that a fast electron on field lines connecting the two stars can be
expected to experience an electric field such that 
${\bf E}\cdot{\bf v}<0$ so that it receives continuing boosts of
energy via equation (\ref{accel}). If the fast electron shuttles back
and forth between mirror points, the maximum length of time that it
spends accelerating rather than decelerating is $\sim2L/c$.  If it is
doomed to strike a polar cap before mirroring, the maximum time is
again $\sim 2L/c$. Integrating equation (\ref{accel}) over the time
interval $2L/c$, we find that the maximum boost yields a Lorentz
$\gamma$ given by 
\be 
\gamma \sim {e \left | {\bf E} \right | 2L/c \over m_e c} 
\sim {e \bint (L/c\tdecay)2L \over m_e c^2} \sim 3
\times 10^5 \, .  
\label{gammaxone} 
\ee 
In fact, such high values of $\gamma$ are inconsistent with limits
obtained from considering the energy drain due to synchrotron losses.
The maximum $\gamma$ consistent with synchrotron losses on a time
scale $\tdecay = t_B$ is given by 
\be 
\gamma_{\rm max} \sim {8 \pi m_{\rm e} c \over \sigma_T \bint^2 t_B} 
\sim 26,000 \, , 
\label{gammaxtwo} 
\ee 
where $\sigma_T= 8\pi r_{\rm e}^2/3 = 6.65 \times 10^{-25} {\rm cm^2}$
is the Thomson cross section with $r_{\rm e} = e^2/m_{\rm e}c^2$ equal
to the classical radius of the electron (see, e.g., Shu 1991).  The
maximum value $\gamma_{\rm max} \sim 26,000$ is considerably larger
than the value $\gamma \sim 127$ required to produce the 90 GHz
synchrotron radiation. Having the maximum value $\gamma_{\rm max}$
larger than the required value of $\gamma$ is not necessarily a
problem, but our arguments below suggest that the synchrotron spectrum
does not extend to frequencies much higher than 90 GHz, so that some
reduction of $\gamma_{\rm max}$ may be indicated.

Toward this end, we speculate that the self-consistent electric fields
are substantially reduced by the collective response of the
intervening semi-relativistic plasma, and that the energies of fast
electrons have to be renewed on each periastron passage because the
particles escape from the system or are lost to brehmsstrahlung as the
mirror points change with reconnecting fields every orbit. The
relevant decay time for the bulk of the fast electrons in the system
is then $\tdecay \sim \porb = 51.1$ d $\sim 4.4 \times 10^6$ s.  The
electrons that can carry over from one outburst to another have a
maximum Lorentz $\gamma_{\rm max} \sim 59$, which implies that the
electrons responsible for the 90 GHz radiation would need
re-generation for each orbit cycle.  A third possibility is that the
component of ${\bf E}$ along magnetic field lines is considerably
smaller than its absolute value $|{\bf E}|$, so that an integral of
equation (\ref{accel}) produces results appreciably smaller than the
naive estimate of equation (\ref{gammaxone}).  We are in the process
of quantitatively evaluating the cumulative result of all three
effects by doing test orbit integrations and will report on the
results in a future communication.

From where then does the time scale $t_B = 10^4$ s arise?  The time
scale for magnetic reconnection is given empirically in the case
of solar flares roughly by the formula:
\be
t_B = {L\over q v_{\rm A}},
\ee
where $v_{\rm A} = B_{\rm int}/(4\pi \rho)^{1/2}$ is the Alfven speed
for a plasma with mass density $\rho$ and $q$ is a dimensionless
number (the Petschek parameter) whose exact value is of considerable
theoretical debate but probably has a value between 0.1 and 1.  In
order to achieve the burst time scale $t_B \sim 10^4$ s with $L =
2\times 10^{12}$ cm, we require an Alfven speed $v_A = 2000 q^{-1}$ km
s$^{-1}$. With our estimate that $\bint = 2$ G, we then deduce a
typical plasma density of $\rho \sim 8.0 \times 10^{-18} q^2$ g
cm$^{-3}$. For a fully ionized gas with solar abundance, such a plasma
would have a number density of electrons given approximately by
$n_{\rm e} \sim 4.2 \times 10^6 q^2$ cm$^{-3}$, well below the value,
for $q \le 1$, that would have caused (an unseen) Faraday
de-polarization of the emitted synchrotron radiation (Phillips et
al. 1996). Note that the communication speed $v_{\rm A}$ is large
compared to the orbital speeds at which the magnetic configuration is
stressed; this ordering justifies our treatment of the background
magnetic evolution as being governed by elliptic partial-differential
equations (the vacuum-field equations).

We still need an explanation for the value of the number density
$n_{\rm e} \sim 4.2 \times 10^6 q^2$ cm$^{-3}$.  The flux density
$S_\nu$ at a frequency $\nu$ = 90 GHz during the radio outburst
reaches a temporal peak (see Figure 2 of Massi et al. 2006) given by
\be
S_\nu = 400 \, {\rm mJy} = 4\times 10^{-24}
\,{\rm erg}\,{\rm s}^{-1}\,{\rm cm}^{-2}\,{\rm Hz}^{-1},
\ee
followed by an exponential decay with mean life 2.3 hr, which we
interpret to be essentially the reconnection time scale $t_B = 10^4$
s, and not the radio-synchrotron decay time scale 
$\tdecay \sim \porb = 51.1$ d.  We compute the total emitted
synchrotron power from the product of the peak $S_\nu$ and $t_B =
10^4$ s integrated over all frequencies.

To perform the integration over $\nu$ we have to specify the energy
distribution of the relativistic electrons. In our situation, the
distribution is established, not by a collisionless shock, but by the
acceleration provided by magnetic reconnection that varies with the
binary orbit balanced by losses through particle removal from the
polar caps (and the X-region; see below). For this problem, we lack
an {\it a priori} theory for the number density $n(\gamma) \, d\gamma$
of fast electrons with Lorentz gamma between $\gamma$ and $\gamma +
d\gamma$.  In Appendix \ref{sec:electrons}, we discuss the case of a
general power-law distribution,
\be
n(\gamma)\, d\gamma = n_0 \gamma^{-p} \, d\gamma \, .
\ee
For the sake of definiteness, we start by specializing to the case
with $p = 2.4$ (the value appropriate for collisionless shocks).  
For $p$ = 2.4, the integral of $n(\gamma) \, d\gamma$ from $\gamma =
1$ to $\gamma = \gamma_{\rm max}$ is insensitive to the cutoff value
$\gamma_{\rm max}$ as long as $\gamma_{\rm max} \gg 1$ (e.g., if
$\gamma_{\rm max} = 127$).  The total number density of fast electrons
in such cases is approximately $n_0/1.4$.

The quantity $n_0$ yields the volume emissivity of synchrotron
radiation as
\be
{C_1}r_{\rm e}n_0eB_\perp \left({\nu \over \nu_\perp}\right)^{-0.7}
\ee
where $r_{\rm e}$ is again the classical radius of the electron,
$\nu_\perp \equiv e B_\perp/2\pi m_{\rm e} c$, and $B_{\perp}$ is the
component of the magnetic field perpendicular to the line of sight
(i.e., in the plane of the sky for angularly resolved observations).
The dimensionless number $C_1\approx4.1$ for our choice $p$ = 2.4,
and is defined in general by equation (\ref{conedef}) in Appendix
\ref{sec:electrons}.  For optically thin synchrotron emission, the
flux density at a distance $d$ from the source is given by
\be
S_\nu = \left[ {1\over 3}
\left( C_1 r_{\rm e} n_0 eB_\perp\right)L\right]
\left({\nu\over \nu_\perp}\right)^{-0.7}\left( {L\over d}\right)^2,
\label{Snu}
\ee
where we have used equation (18.20) of Shu (1991) [which is missing a
factor of $R_s = L$], and where $L$ is the characteristic ``radius''
of the source.

With $S_\nu \propto \nu^{-0.7}$ in equation (\ref{Snu}), the
implied integrated power from zero frequency to $\nu$ is given by
\be
\pnu = 4\pi d^2\int_0^\nu S_\nu \, d\nu = 
4\pi d^2\left(\nu S_\nu/0.3\right) , 
\ee
where $d$ = 133 pc is the distance to V773 Tau (Torres et al. 2011).
If we put in the observed value $S_\nu = 400$ mJy = $4\times 10^{-24}$
erg cm$^{-2}$ s$^{-1}$ Hz$^{-1}$ at $\nu = 90$ GHz, we find that
$\pnu = 2.5 \times 10^{30}$ erg s$^{-1}$ for the synchrotron emission
from zero frequency to $\nu = 90$ GHz.  Over a burst time of $t_B =
10^4$ s, the released energy at radio frequencies up to 90 GHz equals
$2.5\times 10^{34}$ erg, which is nominally 40\% of the total magnetic
energy available in our model in going from apoastron to periastron,
$E_{\rm int}^{\rm apo} - E_{\rm int}^{\rm peri} = 6.1\times 10^{34}
{\rm erg}$ (refer to eq. [\ref{apoperi}]).

The above considerations suggest that if $B_\ast = 1500$ G is a good
estimate for the stellar surface fields, the conversion of magnetic
energy into synchrotron radiation must be efficient, and little
synchrotron radiation is emitted at frequencies higher than 90 GHz.
Because the collisional losses to the thermal plasma are expected to
be negligible (see the discussion in Appendix A of Massi et al.
2006), we believe the acceleration of runaway electrons to be highly
efficient, not only in the sense that a high proportion of eligible
plasma can be driven to relativistic energies, but also to the extent
that there is a limited escape of such accelerated electrons from the
system before they are overtaken by synchrotron losses.  In other
words, as long as the cosmic rays generated within closed field lines
do not open such field lines, almost the entire energy available from
the change of the magnetic configuration will be used to accelerate
the particles responsible for the deduced radio emission. Moreover,
these particles essentially lose all their energy by synchrotron
emission before the next cycle of field dissipation, particle
acceleration, and synchrotron emission; as a result, almost all the
energy input from magnetic field dissipation in the V773 Tau A system
is converted into radio synchrotron emission.  There is, however, a
supra-thermal population of electrons that carries over from one cycle
to another that serves as the seed population (the ``storage ring'')
for the next round of synchrotron acceleration.  In fact, we argue
below that the supra-thermal population may be the predominant
fraction of the plasma density that we estimated using the
reconnection time argument.

For $B_\perp = 2$ G and $\nu/\nu_\perp = (127)^2$, we compute from
equation (\ref{Snu}) and $S_\nu = 400$ mJy at $\nu = 90$ GHz that
$n_0/1.4 = 1.44 \times 10^5$ cm$^3$.  Within the uncertainties of the
calculation, the number density $\nfast$ of fast electrons needed to
account for the observed radio emission is then comparable to the
total electron density $n_{\rm e} = 4.2 \times 10^6 q^2$ cm$^{-3}$
deduced to be present on the basis of the reconnection time. (The
latter calculation was performed non-relativistically, which is
justified for the ions that provide the bulk of the mass density. 
The number density of the electrons, which may be relativistic, is
then deduced on the basis of charge neutrality.) In other words, the
supra-thermal population may be a significant fraction or even all of
the plasma electrons (if $q = 0.19$), which provides the sought-for
explanation for the value of $n_{\rm e}$.  As a bonus, we obtain the
attractive suggestion that the supra-thermal electrons may themselves
be the contributor to the ``anomalous resistivity'' that triggers the
reconnection event. 

That such a source of anomalous resistivity is needed can be seen by
computing the reconnection scale if it occurs by collisional
resistivity: $\ell_R$ = $(t_B \eta)^{1/2} \sim 3 \, {\rm km} \,
(T/10^4 K)^{-3/4}$, where we have used the Spitzer (1978) formula for
the resistivity of a fully-ionized cosmic plasma: $\eta = 10^{13}
T^{-3/2}$ cm$^{2}$ s$^{-1}$. For any plausible value of the
temperature $T$ of a thermal plasma, the length scale $\ell_R$ over
which collisional resistivity can induce reconnection is much smaller
than the scale $L$ over which the spatially-resolved VLBI observations
require the events to occur (Massi et al. 2006).  We conclude that the
true resistivity occurs not by physical collisions knocking charged
particles off field lines, but by their scattering off fluctuating
electromagnetic fields that arise because the laminar collisionless
flow (particularly the counter-streaming of charges driven by
quasi-electrostatic fields from the polar caps of the two stars) is
unstable to the generation of plasma waves.

We can now provide a consistency check on our deduced value of $n_0$
using the following mechanical argument: If we follow convention and
assume that the momentum distribution of cosmic-ray electrons is
isotropic, we can define a cosmic-ray electron pressure $P_{\rm cr}$
through the expression
\be
P_{\rm cr} =
{1\over 3}\int_0^\infty vpn(\gamma)\,d\gamma \approx
{1\over 3}m_{\rm e}c^2 n_0 \int_0^\infty \gamma^{-1.4}
\, d\gamma \approx 0.83 m_{\rm e}c^2 n_0,
\ee
where we have extended the integration to infinity because most of the
contribution comes from low energies, and where we have made the
approximation that the product of speed and momentum, $vp$, for a
relativistic electron equals its energy $\gamma m_{\rm e}c^2$. Using
$n_0/1.4 = 1.44 \times 10^5$ cm$^{-3}$, we find that the electron
pressure $P_{\rm cr} \approx$ 0.14 erg cm$^{-3}$. This value should be
compared with the magnetic pressure $B_{\rm int}^2/8\pi \approx 0.16$
erg cm$^{-3}$.  The pressure of fast electrons at peak outburst is
thus comparable to the typical magnetic field pressure in the
interaction region, implying that the electrons that escape from
beneath the X-point into the region above the X-point in Figure
\ref{fig:blines} can inflate the fields in the latter until they
become open.  Once every orbit cycle, the fast electrons can escape
their confinement through such a mechanism; as a result, the density
$n_0$ cannot accumulate through repeated accelerations to a value that
is much in excess of that calculated here.  Similar considerations in
other contexts may offer an explanation for why the assumption of
rough equipartition of energies in cosmic-ray electrons and magnetic
fields holds in many systems of interest in astrophysics.

For sufficiently low frequencies, synchrotron emission is suppressed
due to the Razin-Tsytovich effect (e.g., Razin 1960, Tsytovich 1951,
Simon 1969, Melrose 1972). The lowest possible value of the frequency
for which waves can propagate is given by
\be
\nu_R = {4 \pi \nu_P^2 \over 3 \omega_L \sin \theta}
\sim 20 {n_e ({\rm cm}^{-3}) \over B ({\rm G}) }{\rm Hz}
\sim 42 \, q^2 {\rm MHz}\,,
\ee
where $\nu_P= ({n_e e^2 / \pi m_e})^{1/2}$, $\omega_L = eB/m_e c$, and
where we have used $n_e = 4.2 \times 10^6 q^2 {\rm cm^{-3}}$,
$B\sim2$G, and $\sin \theta \sim 1$.  This frequency is much lower
than the 90 GHz synchrotron emission observed in the V773 binary
system, so that Razin-Tsytovich suppression does not appear.

We check now whether the modeled synchrotron emission is optically
thin as assumed. The mean synchrotron optical depth $\tau_\nu$ through
the source region is given by
\be
\tau_\nu = {1\over 3} C_2 n_0
\left({cr_{\rm e}\over \nu_\perp}\right)L
\left({\nu\over \nu_\perp}\right)^{-3.2}\,,
\label{tausyn}
\ee
where the constant $C_2\approx9.9$ for $p$ = 2.4, and is defined in
general by equation (\ref{ctwodef}) in Appendix \ref{sec:electrons}.
The right-hand side of equation (\ref{tausyn}) nominally equals
$5\times 10^{-5}$ at 90 GHz, which would indeed make the synchrotron
radiation optically thin at the observed radio frequency. Radio
observations at lower frequencies could be used to measure the
spectral index and to show the bend in the spectrum as a transition is
made to optically thick conditions; such measurements would yield
valuable diagnostic information on the physical conditions in the
system.
 
Finally, we note that the ratio of inverse Compton losses to
synchrotron losses is given by $\ucomp/u_B$ where $\ucomp$ is the
energy density of photons and $u_B = B^2/8\pi$.  Most of the energy
density $\ucomp$ in the V773 Tau A system arises from optical photons
emitted by the two stars; to estimate this contribution, we divide the
optical luminosity of the system, 3.93 $L_\odot$ \citep{boden}, by
$4\pi L^2 c$ to obtain $\ucomp \sim 0.010$ erg cm$^{-3}$.  This value
should be compared with the magnetic counterpart $u_B \sim 0.16$ erg
cm$^{-3}$.  Although inverse Compton losses are significantly smaller
than synchrotron losses, they are not completely negligible.  We
suggest that it might be worthwhile to try to detect optical photons
that have been scattered into the hard X-ray regime by the inverse
Compton effect involving the synchrotron electrons that are deduced
for the V773 Tau A system.

In summary, if our model is to have validity, two conditions must
hold:

\begin{itemize}

\item The synchrotron spectrum $S_\nu$ must not extend to frequencies
much higher than the observed radio frequency of $\nu = 90$ GHz.
At first sight, this requirement seems to be contradicted by the
case of the DQ Tau binary, where an X-ray outburst coincided with
a radio outburst (Getman et al. 2011; see also Salter et al. 2008).  
However, enhanced X-ray emission also occurs in the case of solar
radio outbursts, and the interpretation there is that the fast
particles created by electromagnetic processes in an impulsive flare
or coronal mass ejection travel down the field lines to the footpoints
of a magnetic loop.  The collisions that the fast electrons have with
the solar atmosphere then produce optical and X-ray emission by
brehmsstrahlung radiation.  It would be interesting to know whether
the X-ray outbursts in DQ Tau are synchrotron events (i.e., with a
power-law spectrum and a high degree of linear polarization), or they
are brehmsstrahlung events (i.e., with an optically-thin thermal
spectrum and no associated polarization), or they have the
characteristics of high-energy photons that have been created by
inverse Compton scattering.

\item Although the change of magnetospheric configuration in the
V773 Tau A binary is occurring on the orbital time scale of weeks,
the actual release of the stored energy occurs not on the orbital
time scale (i.e., not as depicted in Fig. 4), but on a much shorter
burst time scale of hours (e.g., $t_B \sim 10^4$ s).  Again this
property is consistent with the solar experience.  Although the
twisting of the coronal magnetic configuration originates in
differential motions at the footpoints of the field in a solar
atmosphere whose rotation rates at the pole and the equator are
measured on a time of many weeks, the outbursts when they come occur
on a time scale of minutes (for impulsive flares) or hours (for
coronal mass ejections).  The build-up of magnetic stresses may
occur slowly and steadily [without instantaneous dissipation as
assumed by the simplifying philosophy adopted at the beginning of
our discussion (see \S 1)], but the relief of such stresses can
appear suddenly and catastrophically as in solar flare events.

\end{itemize}

\subsection{Stellar Winds}

If the stars in the system have substantial stellar winds, their
mechanical luminosity could compete with magnetically produced power
from the interaction region. Scaling the mechanical luminosity 
$\lmech$ of the wind to the outflow rate from the Sun, we find
\be\lmech={1\over2}{\dot M}_w{v_w^2}\approx{5\times10^{26}}\,\ergsec\, 
\left({{\dot M}_w\over10^{-14}\, M_\odot\,\,{\rm yr}^{-1}}\right) 
\left({v_w \over 400 \, {\rm km} \,\, {\rm s}^{-1}}\right)^2.\ee
If the stars in V773 Tau A had solar-type winds, their mechanical
luminosity would be more than 200 times smaller than the magnetic
power scale of equation (\ref{powerscale}).  With a mass-loss rate
$\dot{M}\approx10^{-12}\,{M}_\odot$ yr$^{-1}$ for each star, wind power
would be competitive with magnetic power, and such mass-loss rates are
not beyond the capabilities of weak-line T Tauri stars.

In settings where stellar winds help shape magnetic field structures
in young stars, the configuration could have both dipole and
split-monopole components \citep{mattpud,jardine}, and this form has
been proposed for the solar magnetic field \citep{ban}. The field
strength in a split-monopole declines as $B\propto{r_j}^{-2}$, more
slowly than a magnetic dipole, and collisionless shocks might play a
stronger role in accelerating particles than magnetic
reconnection. However, VLBI mapping for the V773 Tau A system
\citep{massi2008} suggests otherwise; these authors interpret the
magnetic field configurations as solar-like helmut streamers, where
the outflow is not strong enough to overwhelm the fields. As a result,
we suspect that the mass-loss rate in stellar winds in each of these
weak-line T Tauri stars is substantially smaller than
$\dot{M}\approx10^{-12}{M}_\odot$ yr$^{-1}$. Indeed, observations
suggest that typical mass loss rates from weak-lined T Tauri stars are
roughly $\dot{M}\sim10^{-13}M_\odot$ yr$^{-1}$ (Guenther \& Emerson
1997).

Although the mechanical luminosity of the wind could compete with the
magnetic power dissipated in the interaction region, the wind is not
strong enough to greatly alter the magnetic field structure: The 
magnetic pressure in the interaction region is given by $P_B$ =
$B_{\rm int}^2/8\pi$, where we expect $B_{\rm int} \approx$ 2 G, so
that $P_B \approx 0.16$ dyne cm$^{-2}$. For comparison, the ram
pressure of the wind, evaluated at the location of the interaction
region, can be written in the form $P_{\rm ram}$ = 
${\dot M}v_w/(4\pi{r^2})$. For a mass loss rate ${\dot M} = 10^{-12}
M_\odot$ yr$^{-1}$, the ram pressure is only $P_{\rm ram} \approx 5.6
\times 10^{-6}$ dyne cm$^{-2}$, nearly 30,000 times smaller than the
magnetic pressure. Closer to the star, the magnetic pressure dominates
by an even larger margin. As a result, the magnetic field is strong
enough that the wind cannot change the field configuration, and hence
the use of dipole fields remains valid.

\subsection{Back Reaction} 

In this scenario, magnetic energy is released when field lines connect
as the stars move closer together. After periastron, as the stars
separate, energy must be converted back into the magnetic fields in
order to maintain the same dipole strength. This energy could come
from either the stellar orbit or the stellar spins. Using equations
(\ref{eintest}) and (\ref{kepler}), we can estimate the total energy
released per orbit
\be
\Delta{E}\approx\left({2B_\ast^2R_\ast^6\over 3a_0^3}\right)
{\ecc(3+\ecc^2)\over (1-\ecc^2)^3},
\label{totalenergy}
\ee 
where $\asemi$ is the semi-major axis and we have assumed 
$|f_{\rm mean}|\sim 1$.  Using parameters from equation
(\ref{eintnum}), we find the change in energy per orbit
$\Delta{E}\approx 6\times10^{34}$ erg (consistent with
eq. [\ref{apoperi}]). The available mechanical energy in the orbit is
the difference between the energies of the eccentric orbit and the
circular orbit with the same angular momentum, about
$3.5\times10^{45}$ erg. This energy supply can last $6\times10^{10}$
orbits, much longer than the age of the system. The back reaction on
the stellar orbit is thus negligible.

Next we consider the stellar spins, which provide another possible
source of energy for the magnetic interactions. Field lines connecting
the two stars will be twisted by the combined motion of the orbit and
the stellar spins to produce an azimuthal field.  After one rotation
period, these azimuthal field lines must reverse polarity along the
line of centers joining the stars. This magnetic configuration
naturally generates current sheets which in turn trigger reconnection
events (because the field lines cannot be sheared indefinitely),
consistent with the picture developed in this paper. Since the stars
have periods of approximately $\prot \sim 3$ day, both bodies
have total spin kinetic energy $K_{\rm spin} \sim I \omega^2$, where
the moment of inertia $I=k M_\ast R_\ast^2 \sim k(6\times 10^{55})$ g
cm$^2$, and where $k$ is a dimensionless constant.  For example, for a
polytrope with index $n=1.5$ (a good approximation for completely
convective pre-main-sequence stars), one finds $k \approx 0.205$. The
total energy stored in stellar rotation is then given by $K_{\rm spin}
\sim 7 \times 10^{45}$ erg, which is comparable to, but somewhat
larger than, the available energy stored in the orbit. As a result, the
spin energy supply can last for billions of orbits and the
backreaction can (again) be neglected.

\section{Conclusion} 

This paper shows that interactions between the magnetic fields of
young eccentric binaries can provide significant power for
accelerating fast electrons.  The available interaction energy is
large enough to produce the observed radio synchrotron signature in
the well-studied system V773 Tau A.  The energy stored in the magnetic
fields of the system varies with orbital phase, with a maximum rate of
energy dissipation occurring just before periastron (\S 2.3,
Fig. \ref{fig:power}).  This basic model can explain qualitatively the
synchrotron radiation observed in the T Tauri binary V773 Tau A.
However, the details as to how and why the radio flares occur with the
energetics and time scales that they exhibit involve more elaborate
considerations.  These considerations include how magnetic field
annihilation yields electric fields that accelerate electrons to
relativistic energies and how these electrons are ultimately lost to
the system, through synchrotron radiation, or brehmsstrahlung events,
or escape from the system.

In the last regard, we began by assuming that all the field lines in
the system close onto one star or the other.  However, in \S 2.4 we
presented the case that during peak outburst, the fast electrons in
the V773 Tau A system may have enough pressure to open some of the
field lines that join one star to the other.  Through electrostatic
forces, the flow of the electrons to the interstellar medium will
carry a corresponding number of ions with them.  The accelerated fast
electrons that do not escape from the system will ultimately lose
energy, either through collisions with ions (mostly at the polar caps
of the two stars) or through synchrotron radiation. A surprising
prediction of this model for the V773 Tau A system is that the
synchrotron spectrum cuts off at radio frequencies only somewhat
higher than the observed value of 90 GHz. Another prediction is that
the X-ray emission in the young binary DQ Tau (with periastron $\sim
8R_\ast$), which is correlated with the radio bursts \citep{getman},
arises not from synchrotron emission, but probably has a
brehmsstrahlung origin when fast particles associated with the
electromagnetic acceleration stream to the footprints of the magnetic
field and strike the surface of one of the stars in the system.

In closing, we note that the ideas presented in this paper can be
tested.  Perhaps the best way to study these phenomena further is
through observational surveys of synchrotron radiation in a large
sample of pre-main-sequence binary systems. In this model, the
magnetic interaction energy $\eint$ and the idealized dissipated power
$\pmag$ are straightforward functions of the stellar field strengths,
stellar radii, semi-major axis, and orbital eccentricity (eqns.
[\ref{eintest}], [\ref{eintnum}], and [\ref{power}]).  The binary
orbital elements and stellar properties can be measured independently
of the synchrotron radiation. While hysteresis will make the
synchrotron power more sporadic than the idealized dissipation
described by equation (\ref{power}), the total energy release per
orbit should be derivable from equation (\ref{totalenergy}).

\acknowledgments

We thank the Michigan Center for Theoretical Physics for helping to
facilitate this collaboration, and thank Scott Gregory for useful
discussions.  We also thank Ron Taam for suggesting that we check
whether inverse Compton scattering can play a role in V773 Tau A.
This work was supported in part by the Theoretical Institute for
Advanced Research in Astrophysics (TIARA) operated under the Academia
Sinica Institute of Astronomy \& Astrophysics in Taipei, Taiwan (FHS
and MJC), by CONACyT J010/0654/10 (SL), and by NASA Grant NNX11AK87G
(FCA).  Finally, we thank an anonymous referee for many useful
comments that improved the manuscript.


\appendix 

\section{Solid Angle of the Magnetic Polar Cap} 
\label{sec:polarcap} 

As discussed in the text, and shown in Figure \ref{fig:polarcap}, 
the solid angle $A_c$ of the polar cap is given by the approximate
empirical formula $A_c\approx 2R_\ast/L$. This result can be
understood as follows: Half of the circumference of the polar cap
arises from field lines that connect to a straight line of X-points
defined by $(z=0, x = L)$, with $y$ running from $-\infty$ to
$+\infty$.  In spherical polar coordinates $(r,\theta,\varphi)$, the
equation for a magnetic field line of a single dipole placed at the
center of a star (the origin of the coordinate system) is defined by
the equation
\be r = r_0\sin^2\theta, \ee 
where the field line crosses the equatorial plane at $r = r_0$.  The
line of X-points is given by the equation $r_0\cos\varphi = L$ with
$\varphi$ running from $-\pi/2$ to $+\pi/2$.  These lines map onto
the surface of the star $r = R_\ast$ through the equation 
\be 
\sin^2\theta = \left({R_\ast \over L}\right) \cos\varphi,
\label{thetac}
\ee
which describes a semi-curve, $\theta = \theta_c(\varphi)$, at the
polar cap, where the argument varies from $\varphi = -\pi/2$ to
$\varphi = +\pi/2$.  In the expected regime $R_\ast/L \ll 1$, the
contribution $A_X$ of the X-points to the solid angle of the polar 
cap is thus given by 
\be
A_X = \int_{-\pi/2}^{+\pi/2} d\varphi \int_0^{\theta_c(\varphi)}
\sin\theta \, d\theta = \int_{-\pi/2}^{+\pi/2}
\left\{1-\cos\left[\theta_c(\varphi)\right]\right\}\,d\varphi\,.
\label{ax}
\ee
Equation (\ref{thetac}) implies 
\be
\cos[\theta_c(\varphi)] = \left[ 1 - \left( {R_*\over L} \right)
\cos\varphi \right ]^{1/2} \approx 1 - 
\left({R_\ast \over 2L}\right) \cos\varphi .
\ee
Using this expression in equation (\ref{ax}) 
yields the approximation 
\be
A_X \approx {R_\ast \over L}.
\ee
From Figure \ref{fig:polarcap}, we find that the other semi-curve
defined by the magnetic field loops that connect one star to the other
via $z = +\infty$ contributes (approximately) an equal amount. The
solid angle of the polar cap thus becomes 
\be 
A_c \approx { 2 R_\ast \over L},
\ee
in agreement with the formula found empirically.

\section{Generalized Electron Energy Distribution} 
\label{sec:electrons}

In this Appendix, we generalize the calculation of \S 2.4 to include a
general power-law form for the energy distribution of fast electrons,
i.e., 
\be
n(\gamma)\,d\gamma = n_0\gamma^{-p} \,d\gamma ,
\ee
where the index $p$ lies in the range $1 \le p \le 2.4$. The lower
limit leads to a flat spectral energy distribution $\nu S_\nu \sim$
{\it constant}, whereas the upper limit is appropriate for
collisionless shocks.

The number density $n_T$ of fast particles, integrated from $\gamma=1$
to $\gamma=\infty$, is finite, and is given by $n_T=n_0/(p-1)$ when
$p>1$.  However, the integrated electron pressure is given by 
\be
P_{\rm cr} = {1\over 3}\int_1^\infty 
v\gamma m_{\rm e}v n(\gamma)\, d\gamma,
\ee
which diverges unless $p>2$. For $p<2$, one must introduce an upper
cuttoff in $\gamma$ to prevent the divergence of the pressure. 
For the case $p>2$, with $v \approx c$, the pressure  
$P_{\rm cr} \approx n_0 m_{\rm e}c^2/[3(p-2)]$. 

The volume emissivity of synchrotron radiation takes the 
generalized form 
\be
C_1 r_{\rm e}n_0eB_\perp \left({\nu \over \nu_\perp}\right)^{-(p-1)/2},
\ee
where the dimensionless number $C_1$ is defined by 
\be
C_1 = \left( {3^{p/2} 2 \over p+1}\right)
\Gamma\left({p\over 4} - {1\over12}\right)
\Gamma\left({p\over 4} +{19\over 12}\right),
\label{conedef} 
\ee
where $\Gamma(\zeta)$ is the Gamma function.  For optically thin
synchrotron emission, the flux density at a distance $d$ from the
source is given by
\be
S_\nu = \left[ {1\over 3}\left( C_1 r_{\rm e} n_0 eB_\perp\right)L\right]
\left({\nu\over \nu_\perp}\right)^{-(p-1)/2}\left( {L\over d}\right)^2.
\label{snupnew}
\ee
With $S_\nu \propto \nu^{-(p-1)/2}$ in equation (\ref{snupnew}), the 
implied power $\pnu$, integrated from zero frequency to $\nu$, 
is given by
\be
\pnu = 4\pi d^2\int_0^\nu S_\nu \, d\nu = 
4\pi d^2\left({2\nu S_\nu \over 3-p}\right).
\ee
For the system parameters of V773 Tau A, we find the power $\pnu$ =
$(1.53\times 10^{30})/(3-p)$ erg s$^{-1}$ for synchrotron emission
integrated from zero frequency to $\nu = 90$ GHz. Over a burst time
$t_B=10^4$ s, the released energy $E\approx(1.53\times{10}^{30})/(3-p)$ 
at radio frequencies up to 90 GHz.

Using $B_\perp = 2$ G, $\nu/\nu_\perp = (127)^2$, and $S_\nu = 400$
mJy at 90 GHz, we find that the total number $n_T$ of fast electrons
is given by
\be
n_T = {n_0\over p-1} = {939\over(p-1) C_1} (127)^{(p-1)} \rm{cm}^{-3}.
\ee
Table 1 presents the number density $n_T$ and the parameter $C_1$ for
a range of indices $p$.  Note that for $2< p<2.4$, the density of fast
electrons is consistent with the requirements arising from the
reconnection-time argument (see text).

\begin{deluxetable}{lllll}
\setlength{\tabcolsep}{0.2in}
\tablecolumns{5}
\tablewidth{0pc}
\tablecaption{Electron Density and 90 GHz Optical Depth}
\tablehead{
\colhead{$p$}&\colhead{$C_1$}&\colhead{$C_2$}
&\colhead{$n_T\,\,[{\rm cm}^{-3}]$}&\colhead{$\tau_\nu$} \\
}
\startdata
1.1  & 7.94 & 6.68 & $1.92 \times 10^3$ & $2.41 \times 10^{-4}$ \\
1.5  & 5.52 & 7.17 & $3.83 \times 10^3$ & $7.44 \times 10^{-5}$ \\
2.0  & 4.42 & 8.38 & $2.70 \times 10^4$ & $5.44 \times 10^{-5}$ \\
2.4  & 4.12 & 9.90 & $1.44 \times 10^5$ & $4.92 \times 10^{-5}$ \\
\enddata
\end{deluxetable}

The mean synchrotron optical depth through the source
region at frequency $\nu$ is given by
\be
\tau_\nu = {1\over 3} C_2 n_0 
\left({cr_{\rm e}\over \nu_\perp}\right)L
\left({\nu\over \nu_\perp}\right)^{-(p+4)/2},
\ee
where
\be
C_2 = 3^{(p+1)/2}\Gamma\left({3p+2\over 12}\right)
\Gamma\left({3p+22\over 12}\right).
\label{ctwodef} 
\ee
Table 1 also presents the optical depth $\tau_\nu$ (at $\nu$ = 90 GHz)
and the parameter $C_2$ for varying indices $p$.  For all plausible
energy distributions of the fast electrons, those with $p>1$, the
synchrotron radiation is optically thin at the observed radio
frequency $\nu$ = 90 GHz.  Note that radio observations at lower
frequencies will provide important additional information regarding
the physical conditions in the system (see text).


\end{document}